\documentstyle[extras,epsfig,citews,twoside]{article}

\newcommand{\bcn}{\begin{center}}
\newcommand{\beq}{\begin{equation}}
\newcommand{\beqn}{\begin{eqnarray}}
\newcommand{\ecn}{\end{center}}
\newcommand{\eeq}{\end{equation}}
\newcommand{\eeqn}{\end{eqnarray}}
\newcommand{\case}[2]{\mbox{\small $\displaystyle \frac{#1}{#2}$}}
\newcommand{\dcsb}{\mbox{D$\chi$SB}}

\newcommand{\Eq}[1]{Eq.~(\ref{#1})}

\newcommand{\qbq}{\mbox{$\langle\overline{q}q\rangle$}}

 \def\lsim{\mathrel{\rlap{\lower4pt\hbox{\hskip1pt$\sim$}}
    \raise1pt\hbox{$<$}}}         
\begin{document}
\rightline{ANL Preprint Number: PHY-7854-TH-94}
\hspace*{-\parindent}{\Large\bf \footnote[2]{Summary of a presentation at
{\it Chiral Dynamics: Theory and Experiment}, Cambridge, MA, 25-29 July,
1994.}Pion Observables and QCD
}\vspace*{1cm}

\hspace*{-\parindent}Craig D. Roberts\vspace*{\lineskip}

\hspace*{-\parindent}Physics Division, Bldg. 203, Argonne National
Laboratory,\\
        Argonne, IL 60439-4843, USA

\vspace*{1cm}


\subsection{Introduction}
\markboth{Craig D. Roberts}{Pion Observables and QCD}
\setcounter{equation}{0}
The Dyson-Schwinger equations (DSEs) are a tower of coupled integral
equations that relate the Green functions of QCD to one another.  Solving
these equations provides the solution of QCD.  This tower of equations
includes the equation for the quark self-energy, which is the analogue of the
{\it gap equation} in superconductivity, and the Bethe-Salpeter equation, the
solution of which is the quark-antiquark bound state amplitude in QCD.  The
application of this approach to solving Abelian and non-Abelian gauge
theories is reviewed in Ref.~\citenum{DSErev}.

The nonperturbative DSE approach is being developed as both: 1) a
computationally less intensive alternative and; 2) a complement to numerical
simulations of the lattice action of QCD.  In recent years, significant
progress has been made with the DSE approach so that it is now possible to
make sensible and direct comparisons between quantities calculated using this
approach and the results of numerical simulations of Abelian gauge
theories.\cite{QEDCJB}

Herein the application of the DSE approach to the calculation of pion
observables is described\cite{CDRpion} using: the $\pi$-$\pi$ scattering
lengths ($a_0^0$, $a_0^2$, $a_1^1$, $a_2^0$, $a_2^2$) and associated partial
wave amplitudes; the \mbox{$\pi^0\rightarrow \gamma\gamma$} decay width; and
the charged pion form factor, $F_\pi(q^2)$, as illustrative examples.  Since
this approach provides a straightforward, microscopic description of
dynamical chiral symmetry breaking (\dcsb ) and confinement, the calculation
of pion observables is a simple and elegant illustrative example of its power
and efficacy.  The relevant DSEs are discussed in Sec.~\ref{sec_two}, the
calculation of pion observables in Sec.~\ref{sec_three} and concluding
remarks are presented in Sec.~\ref{sec_four}.

\subsection{Dyson-Schwinger Equations in QCD}
\label{sec_two}
\setcounter{equation}{0}
\subsubsection{Quark Propagator}
In Euclidean space, with metric \mbox{$\delta_{\mu\nu} = \;{\rm diag}
(1,1,1,1)$} and $\gamma_\mu = \gamma_\mu^\dagger$, and in a general covariant
gauge, the inverse of the dressed quark propagator can be written as
\begin{equation}
 S^{-1}(p) = i\, \gamma\cdot p\, + m + \Sigma(p)
	   \equiv Z^{-1}(p^2) \left(i\, \gamma\cdot p + M(p^2) \right)
           \equiv i\, \gamma\cdot p \,A(p^2)\,+ m + B(p^2)\/,
\end{equation}
with: $m$\ the renormalised, explicit chiral symmetry breaking mass (if
present); $\Sigma(p)$\ the self-energy; $M(p^2) = [m + B(p^2)]/A(p^2)$\ the
dynamical quark mass function; and $Z(p^2) = A^{-1}(p^2)$ the
momentum-dependent renormalisation of the quark wavefunction.  The DSE for
the inverse propagator is
\begin{equation}
 S^{-1}(p) = i\,\gamma\cdot p + m +
     \case{4}{3} g^2 \int \frac{d^4k}{(2\pi)^4} \gamma_\mu
         S(k) \Gamma_\nu (k,p) D_{\mu \nu}((p-k)^2)\/,     \label{fullDSE}
\end{equation}
where $D_{\mu\nu}(q^2)$\ is the dressed gluon propagator and $\Gamma_\nu$ is
the proper quark-gluon vertex.

When the current-quark mass is zero, the solution of this equation determines
whether or not chiral symmetry is dynamically broken in QCD.  The quark
condensate, $\qbq \propto \mbox{tr$[S(x=0)]$}$, is a chiral symmetry order
parameter.  If, with $m=0$ in \Eq{fullDSE}, there is a solution with $B\neq
0$ then the quark has generated a mass via interaction with its own gluon
field and the chiral symmetry is therefore dynamically broken.

The solution also provides information about quark confinement.  The presence
or absence of quark production thresholds in the $\cal S$-matrix amplitudes
that contribute to physical observables is determined by the analytic
structure of the quark propagator, which one obtains by solving this
equation.\cite{RWK92}

\subsubsection{Gluon Propagator}

In a general covariant gauge the dressed gluon propagator can be written:
\begin{equation}
\label{Gprop}
D^{\mu\nu}(q^2) = \left[ \left(\delta^{\mu\nu}-\frac{q^\mu q^\nu}{q^2}\right)
  \frac{1}{1-\Pi(q^2)} + \xi \frac{q^\mu q^\nu}{q^2} \right]
\frac{1}{q^2}\/,
\end{equation}
where $\Pi(q^2)$\ is the gluon vacuum polarisation and $\xi$\ is the gauge
parameter.

\begin{figure}[htb] 
 \centering{\ \epsfig{figure=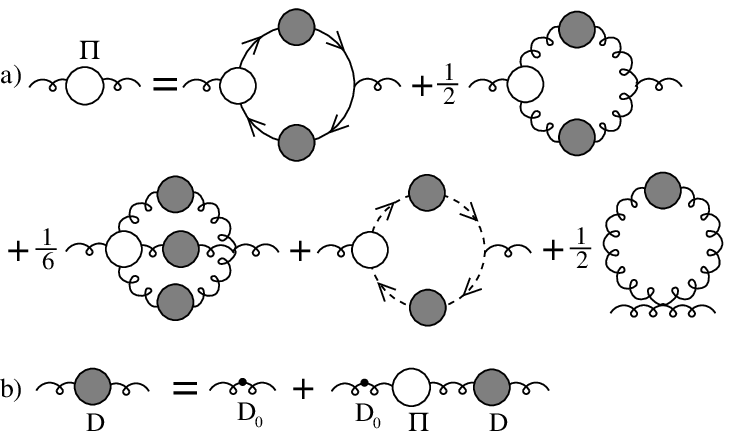,rheight=5cm,height=5.0cm} }
\parbox{122mm}{
\caption{ The Dyson-Schwinger equation for the gluon propagator.
[Here and below the broken line represents the
propagator for the ghost field.]
\label{gluon_dse_fig}  }}
\end{figure}

The Dyson-Schwinger equation for the gluon propagator is given
diagrammatically in Fig.~\ref{gluon_dse_fig}.
The symmetrisation factors of 1/2 and 1/6 arise from the usual Feynman rules,
which also require a negative sign [unshown] to be included for every fermion
and ghost loop.  This equation has been studied
extensively.\cite{SM79,UBG80,BBZBP,H90b} There have also been attempts to
determine the gluon propagator from numerical simulations of
lattice-QCD.\cite{MO,BPS94}

\begin{figure}[htb] 
\psrotatefirst \centering{\
\epsfig{figure=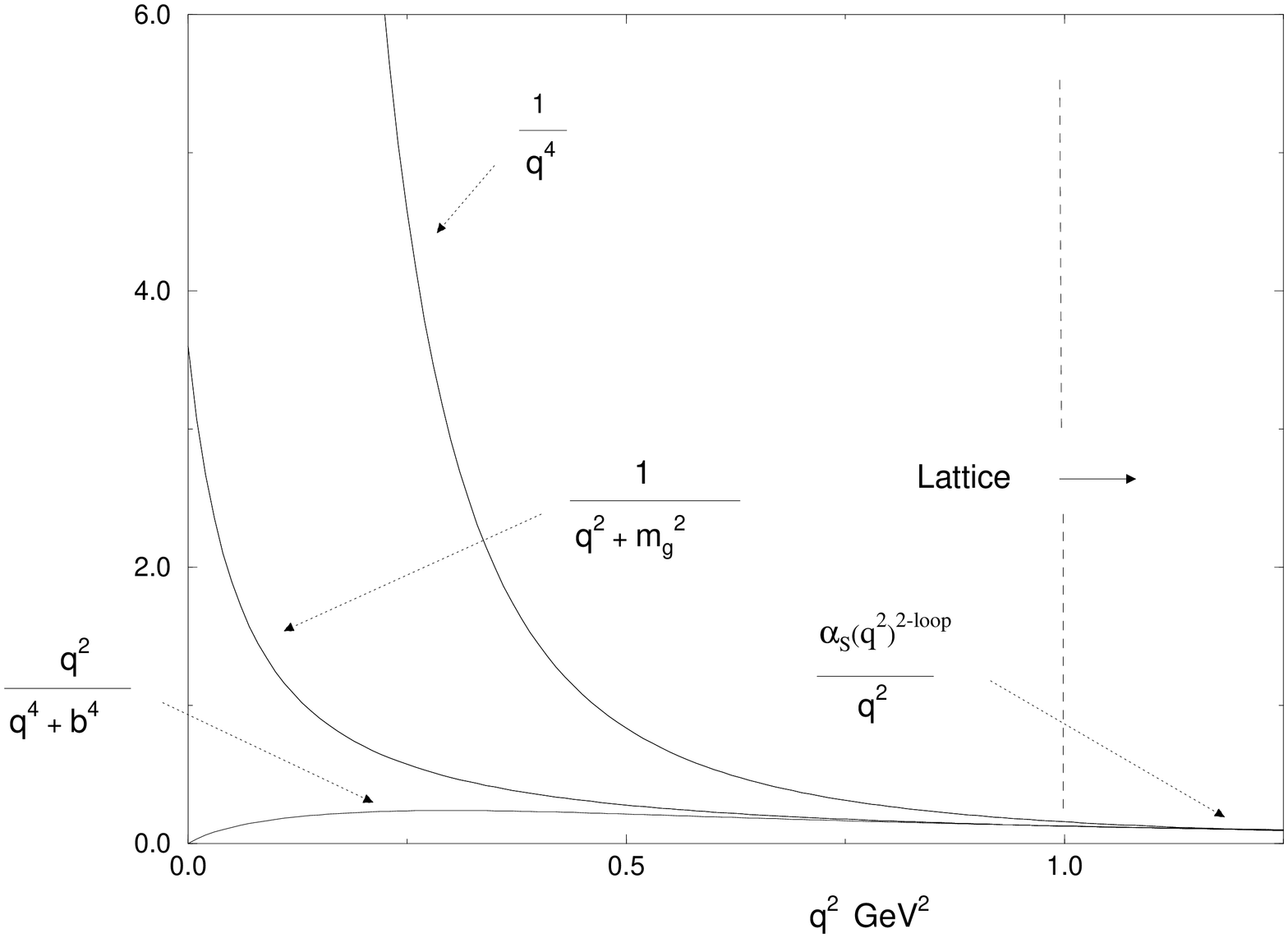,rheight=8.0cm,height=10.5cm,angle= -90} }
\parbox{122mm}{
\caption{Results of studies of the gluon propagator in QCD.  Typical values
of $m_g = 0.5$~GeV and $b= 0.7$~GeV have been used.
\label{plot_D} } }
\end{figure}

The results of the DSE and lattice studies are summarised in Sec.~5.1 of
Ref.~\citenum{DSErev} and are represented in Fig.~\ref{plot_D}.
This figure illustrates that for spacelike-$q^2> 1$~GeV$^2$ the gluon
propagator is given by the two-loop, QCD renormalisation group result, with
the next order correction being $<$10\%.  For spacelike-$q^2< 1$~GeV$^2$,
however, the form of the propagator is not known.

The DSE studies of Refs.~\citenum{SM79,UBG80,BBZBP} suggest a regularised
infrared singularity, represented by $1/q^4$ in the figure. That of
Ref.~\citenum{H90b}, which differs mainly in that the Ansatz used for the
triple-gluon vertex has kinematic singularities, suggests an infrared
vanishing form, characterised by $q^2/(q^4+b^4)$, which has also been
argued\cite{VG79Zw91} to be the form necessary to completely eliminate Gribov
copies.

The lattice Landau-gauge simulations of Ref.~\citenum{MO} favour the massive
vector boson form, $1/(q^2+m_g^2)$, which is broadly consistent with the
improved simulations of Ref.~\citenum{BPS94}.  In some cases, however, these
improved simulations allowed a fit of the form $q^2/(q^4+b^4)$, with $b\sim
340$~MeV.  There is a problem with lattice simulations, however, indicated by
the dashed vertical line at the right of Fig.~\ref{plot_D}.  With present
technology, the domain of $q^2< 1$~GeV$^2$ is actually inaccessible in
lattice studies and, since all forms of the propagator are the same outside
this domain, it is clear that these results are both qualitatively and
quantitatively unreliable.

%
%

\subsubsection{Bound States: Bethe-Salpeter Equation}
In quantum field theory, the Bethe-Salpeter amplitude for a two-body,
quark-anti\-quark bound state is obtained as the solution of the homogeneous
Bethe-Salpeter equation:
\begin{equation}
\label{piBSE}
\Gamma^{rs}_M(p;P) = -
\int\,\case{d^4k}{(2\pi)^4} \,K^{rs;tu}_{ab;cd}(k,p;P)\,
\left(S(k-\case{1}{2}P)\Gamma_M(k;P)S(k+\case{1}{2}P)\right)^{tu}_{cd},
\end{equation}
where $\Gamma_M$ is the proper meson-quark vertex and $K$ is the kernel.  A
commonly used approximation for the kernel, $K$, is {\it generalised ladder
approximation}, in which
\beq
K = g^2 \, D_{\mu\nu}(p-q)\, \left(\frac{\lambda^a}{2}\right)_{ac}
        \left(\frac{\lambda^a}{2}\right)_{db}
        \left(\gamma_\mu\right)_{rt}
                \left(\gamma_\mu\right)_{us}
\eeq
and $S(k)$ is obtained from the {\it rainbow approximation} to the quark DSE;
i.e., \Eq{fullDSE} with $\Gamma_\mu = \gamma_\mu$.  Solving for $\Gamma_M$ in
this approximation yields what is meant by the {\it dressed-quark core} of
the bound state.

In the DSE approach the dichotomy of the pion as both a Goldstone boson and a
\mbox{$q$-$\bar q$} bound state is beautifully and easily understood.  One has
\dcsb\ when, with $m=0$ in \Eq{fullDSE}, one obtains
\mbox{$B_{m=0}(p^2)\neq 0$} as a solution.  In this case $(m=0)$ the
Bethe-Salpeter equation in the pseudoscalar channel reduces to the quark DSE
as $P^2\rightarrow 0$, where $P_\mu$ is the centre-of-mass momentum of the
bound state.\cite{DS79} It follows {\it without fine-tuning}, therefore, that
if the quark DSE admits \mbox{$B_{m=0}(p^2)\neq 0$} as a solution when $m=0$;
i.e., one has \dcsb, then there exists a massless $(P^2=0)$, pseudoscalar,
$q$-$\bar q$ bound state with the proper meson-quark vertex dominated by its
$\gamma_5$ component; i.e.,
\mbox{$\Gamma_\pi(p^2;P^2=0) = B_{m=0}(p^2)/f_\pi$}, where $f_\pi$ is both
the Bethe-Salpeter normalisation and pion decay constant, which is a {\it
calculated} quantity in this approach.  For $m\neq 0$ but $m/\Lambda_{\rm
QCD}
\ll 1$ one has
\beqn
\label{Gpi}
\Gamma_\pi(p^2;P^2= - m_\pi^2) & \approx & \frac{1}{f_\pi}B_{m=0}(p^2)\\
\label{mpi}
m_\pi^2 f_\pi^2 & \approx& 2 m \langle \bar q q \rangle.
\eeqn

\subsubsection{Solution of the Quark Dyson-Schwinger Equation}
The DSE for the quark propagator, \Eq{fullDSE}, has been much studied (see
Sec.~6 of Ref.~\citenum{DSErev}).  The numerical solution obtained using a
gluon propagator with an integrable singularity in the infrared region (upper
curve in Fig.~\ref{plot_D}) and a quark-gluon vertex of the Ball-Chiu
type\cite{BC80} can be represented well by the following algebraic
parametrisation:\cite{CDRpion}
\beqn
\label{SSM}
\lefteqn{\bar\sigma_S(x)  =  C e^{-2 x} +}\\
& & \frac{1 - e^{- b_1 x}}{b_1 x}\,\frac{1 - e^{- b_3 x}}{b_3 x}\,
        \left( b_0 + b_2 \frac{1 - e^{- \Lambda x}}{\Lambda\,x}\right)
        + \frac{\bar m}{x + \bar m^2}
                \left( 1 - e^{- 2\,(x + \bar m^2)} \right) \nonumber
\eeqn
\beq
\label{SVM}
\bar\sigma_V(x)  =  \frac{2 (x+\bar m^2) -1
                + e^{-2 (x+\bar m^2)}}{2 (x+\bar m^2)^2}
                - \bar m C e^{-2 x},
\eeq
where $S(p)= -i\gamma\cdot p \,\sigma_V(p^2) + \sigma_S(p^2)$, $x=p^2/(2 D)$,
$\bar\sigma_S(x) = \sqrt{2 D}\sigma_S(p^2)$, $\bar\sigma_V(x) = (2
D)\sigma_V(p^2)$, $\bar m = m/\sqrt{2 D}$, and $D$ sets the mass scale.  This
parametrisation provides a representation of the quark propagator that is an
entire function in the complex $p^2$ plane but for an essential singularity,
as suggested by Ref.~\citenum{BRW92}, which is sufficient to ensure
confinement.

Equations (\ref{SSM}) and (\ref{SVM}) provide a six parameter algebraic
approximation to the quark propagator in QCD: $C$, $\bar m$, $b_0, \ldots,
b_3$.  [$\Lambda (= 10^{-4})$ simply decouples $b_2$ from the quark
condensate.]  The parameters can be chosen so as to provide an accurate
approximation to a numerical solution of \Eq{fullDSE}.  However, as
illustrated in Fig.~\ref{plot_D}, the form of the gluon propagator for
$q^2<1$~GeV$^2$ is unknown.  This parametrisation is therefore also an
implicit parametrisation of the gluon propagator in this region.  Using this
quark propagator to calculate experimental observables, and choosing the
parameters so as to obtain the best possible fit to these observables, one
has a direct connection between observables and the form of the effective
quark-quark interaction in the infrared.  This provides a means for using
precise, low-energy experimental data to determine the effective quark-quark
interaction in the infrared.

\subsection{Calculating Pion Observables}
\label{sec_three}
\setcounter{equation}{0}
In much the same way as an effective action can be used to formalise the
constraints of chiral Ward identities in QCD, one can formalise the {\it
Abelian approximation} to QCD (which decouples the ghost fields and reduces
the Slavnov-Taylor identities to Ward identities; see Ref.~\citenum{UBG80}
and Sec.~5 of Ref.~\citenum{DSErev}) via a model field theory described by
the action:\cite{CR85}
\beq
S[\bar q,q] = \int\,d^{4}x\,d^{4}y\,
\left\{\bar q(x)\left[\gamma\cdot\partial+m\right]\delta^{4}(x-y)q(y)
+\frac{1}{2}j^{a}_{\mu}(x)g^{2}D_{\mu\nu}(x-y)j_{\nu}^{a}(y)\right\}
\eeq
with
\mbox{$j_{\mu}^{a}(x) =
\bar q(x)\,\left(\lambda^{a}/2\right)\,\gamma_{\mu}q(x)$}
and $D_{\mu\nu}$ the gluon propagator.  This model field theory is called the
Global Colour-symmetry Model (GCM).

\subsubsection{$\pi$-$\pi$ Scattering}
The $\pi$-$\pi$ scattering $T$-matrix is completely specified by one scalar
function, $A(s,t,u)$:
\beq
T_{\alpha\beta;\gamma\delta} =
A(s,t,u)\,\delta_{\alpha\beta}\delta_{\gamma\delta}
+ A(t,s,u)\, \delta_{\alpha\gamma}\delta_{\beta\delta}
+ A(u,t,s)\, \delta_{\alpha\delta}\delta_{\beta\gamma}~.
\eeq
In the DSE-GCM approach the {\it tree-level} contribution to $A(s,t,u)$ is
obtained from a sum of {\it intrinsically-finite} quark-loop diagrams,
which, using Eqs.~(\ref{SSM}) and (\ref{SVM}), have no quark production
thresholds.  Although it is not necessary, one can perform a derivative
expansion of this sum of quark-loop contributions to obtain:\cite{RCSI94}
\beqn
\label{Astu}
\lefteqn{A(s,t,u)  = \frac{ m_{\pi}^2 + 2\,s - t - u}{3\,f_\pi^2}
        + \frac{4 N_c}{3 f_{\pi}^{4}} \times }\\
& & \left[\rule{0mm}{4mm}
K_1 \left( -12\,{m_{\pi}^4} +
6\,{m_{\pi}^2}\,(s + t +u) + 2\,{s^2} - {t^2}- {u^2}
- 2(\,s\,t +\,s\,u +\,t\,u ) \right)\right. \nonumber \\
& &
+ K_2 \left.
\left( -2\,{m_{\pi}^2} + s \right) \,\left( -2\,{m_{\pi}^2} + t + u \right)
+ K_3
\left(-2\,{m_{\pi}^4} + {m_{\pi}^2}\,(s +t+u)  -t\,u\right)
\rule{0mm}{4mm}\right], \nonumber \\
\label{Fpi}
\lefteqn{\mbox{where}\;f_{\pi}^2   =
\frac{N_c}{8\pi^2}\int_{0}^\infty\,ds\,s\,B_{m\neq 0}^2 \,\times} \\
& & \left(  \sigma_{V}^2 -
2 \left[\sigma_S\sigma_S' + s \sigma_{V}\sigma_{V}'\right]
- s \left[\sigma_S\sigma_S''- \left(\sigma_S'\right)^2\right]
- s^2 \left[\sigma_V\sigma_V''- \left(\sigma_V'\right)^2\right]\right)~,
\nonumber
\eeqn
and $m_\pi$ is obtained from \Eq{mpi} with
\beqn
\label{qbarq}
\lefteqn{-\langle \bar q q \rangle_\mu = }\\
& &
\left(\ln\frac{\mu^2}{\Lambda_{QCD}^2}\right)^{\alpha}
\lim_{\Lambda_{UV}^2\rightarrow \infty}
\left(\ln\frac{\Lambda_{UV}^2}{\Lambda_{QCD}^2}\right)^{-\alpha}
\frac{3}{4\pi^2}\int_0^{\Lambda_{UV}^2}\,ds\,s\left( \sigma_S(s)
        - \frac{m}{s+m^2}\right)~. \nonumber
\eeqn
The constants $K_i$ are also given by one dimensional integrals whose
integrands involve only the quark propagator.\cite{RCSI94}  Clearly, and
importantly, each of the quantities appearing in \Eq{Astu} is determined once
the quark propagator is specified.  The utility of the derivative expansion
is only that it facilitates a comparison with other calculations of
$A(s,t,u)$.

\subsubsection{$\pi^0\rightarrow \gamma\gamma$}
In Euclidean metric the matrix element for the decay $\pi^0\rightarrow
\gamma\gamma$ can be written
\begin{equation}
\label{Mpi0}
{\cal M}(k_1,k_2)= -2\,i\, \frac{\alpha_{em}}{\pi f_\pi}\,
        \epsilon_{\mu\nu\rho\sigma}
        \epsilon_\mu(k_1)\,\epsilon_\nu(k_2)\,k_{1\rho}\,k_{2\sigma}\,
                G(k_1\cdot k_2)~,
\end{equation}
where $k_i$ are the photon momenta and $\epsilon(k_i)$ are their polarisation
vectors.  Here, the $\pi^0$ momentum is $P=(k_1 + k_2)$ and $P^2=2\,k_1\cdot
k_2$.

Using Eq.~(\ref{Mpi0}) one finds easily that
\mbox{$\Gamma_{\pi^0\rightarrow \gamma\gamma} =
[m_\pi^3\alpha_{em}^2 G(-m_\pi^2)^2]/[16\pi^3 f_\pi^2]$}.
%
Experimentally one has \mbox{$
\Gamma_{\pi^0\rightarrow \gamma\gamma} = (7.74 \pm 0.56)~\mbox{eV}
$}, which corresponds to
\begin{equation}
g_{\pi^0\gamma\gamma} \equiv G(-m_\pi^2) = 0.504 \pm 0.019~,
\end{equation}
using $m_{\pi^0}= 135$~MeV and $f_\pi=93.1$~MeV.

In generalised impulse approximation $g_{\pi^0\gamma\gamma}$ is obtained from
the sum of two quark-loop diagrams and, in the chiral limit $(P^2 = 0)$, one
has:\cite{CDRpion}
\begin{eqnarray}
\label{G0}
\lefteqn{g_{\pi^0\gamma\gamma}^0 \equiv G(0) =
\int_0^\infty\,ds\,s\,B_{m = 0}\,A\,\sigma_V\,\times} \\
& & \left\{ \rule{0mm}{4mm}
A\,\left[\sigma_V\,\sigma_S
        + s \left(\sigma_V'\,\sigma_S - \sigma_V\,\sigma_S'\right)\right]
+ s\,\sigma_V\,\left(A'\,\sigma_S - B'\sigma_V\right)\right\}~.
\nonumber
\end{eqnarray}
Defining \mbox{$C(s) = B_{m = 0}(s)^2/[s\,A(s)^2]$} one obtains a dramatic
simplification and, because of \dcsb; i.e., because $f_\pi\Gamma_\pi =
B_{m=0}$,
\begin{equation}
\label{gpigg0}
g_{\pi^0\gamma\gamma}^0 = \int_0^\infty\,dC\,\frac{1}{(1+C)^3} = \frac{1}{2}~,
\end{equation}
since $C(s=0)=\infty$ and $C(s=\infty)=0$.  Hence, the experimental value is
reproduced, {\it independent} of the details of $S(p)$.  This illustrates the
manner in which the Abelian anomaly is incorporated in the DSE framework.
This result will be violated in any approach that does not properly
incorporate \dcsb; i.e., in any approach which violates the chiral-limit
identity: $f_\pi\,\Gamma_\pi = B_{m = 0}$.

\subsubsection{$F_\pi(q^2)$}
In generalised impulse approximation, in Euclidean metric, with $\gamma_\mu$
hermitian, the $\pi$-$\pi$-$\gamma$ vertex is\cite{CDRpion}
\begin{eqnarray}
\label{LFpi}
\lefteqn{\Lambda_\mu(P+q,-P)=\frac{N_c}{f_\pi^2}\,
\int\case{d^4k}{(2\pi)^4}\,} \\
& &  {\rm tr}_D
\left[ i\overline{\Gamma}_\pi(k;P+q)
        S(k_{++})i\Gamma_\mu^{\rm BC}(k_{++},k_{-+})S(k_{-+})
        i\Gamma_\pi(k-\case{1}{2}q;-P) S(k_{--})\right],
\nonumber
\end{eqnarray}
where $q$ is the photon momentum, $P$ is the initial momentum of the
pion,\linebreak
\mbox{$k_{\alpha\beta} = k + \case{\alpha}{2} q + \case{\beta}{2}P$}
and\cite{BC80}
\beq
\Gamma_\mu^{\rm BC}(p,q)  =  \Sigma_A(p,q)\,\gamma_\mu
        + (p+q)_{\mu}\left\{ \Delta_A(p,q)\,\case{1}{2}\,
                \left[ \gamma\cdot p + \gamma\cdot q\right]
- i\Delta_B(p,q)\right\}
\eeq
with $\Sigma_H(p,q) \equiv[H(p^2) + H(q^2)]/2$ and $\Delta_H(p,q) \equiv
[H(p^2) - H(q^2)]/[p^2 - q^2]$, for $H=A$ or $B$. $\Gamma_\mu^{\rm BC}$
satisfies the Ward-Takahashi identity and ensures the conservation of the
pion current\cite{CDRpion} so that
\mbox{$\Lambda_\mu(P+q,-P)$}$=(2P_\mu+q_\mu)F_\pi(q^2)$.

\subsubsection{Fitting the Parameters and Calculated Results}
The parameters in Eqs.~(\ref{SSM}) and (\ref{SVM}) are fixed by requiring a
global best-fit to:
\begin{equation}
\begin{array}{ccc}
\displaystyle
\frac{f_\pi}
{\langle \overline{q}q\rangle^{\frac{1}{3}}} = 0.423 \pm 0.024~, &\;\;
f_\pi\,r_\pi = 0.318 \pm 0.006~,\;\;
\displaystyle
\frac{m_\pi^2}
{\langle \overline{q}q\rangle^{\frac{2}{3}}} = 0.396 \pm 0.036~;
\end{array}
\end{equation}
the dimensionless $\pi$-$\pi$ scattering lengths
\begin{equation}
\begin{array}{ll}
a_0^0 = 0.21 \pm 0.02~,  & a_0^2 = -0.040 \pm 0.003~, \\
a_1^1 = 0.038 \pm 0.002~, & a_2^0 = 0.0017 \pm 0.0003~;
\end{array}
\end{equation}
and a least-squares fit to $F_\pi(q^2)$ on the spacelike-$q^2$ domain:
$[0,4]$ GeV$^2$.  The fitting procedure used $f_\pi$ from \Eq{Fpi} and
$\langle\bar q q\rangle$ from \Eq{qbarq}, with $\Lambda_{QCD} = 0.20$~GeV and
$\alpha=1$ [not the true anomalous dimension because $\ln[p^2]$-corrections
have not been included in Eqs.~(\ref{SSM}) and (\ref{SVM})] and the
expressions for $a_0^0$, $a_0^2$, $a_1^1$, $a_2^0$ and $r_\pi$ given in
Ref.~\citenum{RCSI94}.  It yielded
\begin{eqnarray}
& & \begin{array}{cc}
C = 0.0406~, & \overline{m} = 0.0119~,
\end{array} \nonumber \\
& & \begin{array}{cccc}
b_0 = 0.118~, & b_1 = 2.51~, & b_2 = 0.525 ~,& b_3 = 0.169~.
\end{array}
\label{ParamV}
\end{eqnarray}
The mass scale is set by requiring equality between the percentage error in
$f_\pi$ and $r_\pi$, which yields $D=0.133$~GeV$^2$.

\begin{figure}[htb] 
\vspace*{-4.5cm}
 \centering{\
  \epsfig{figure=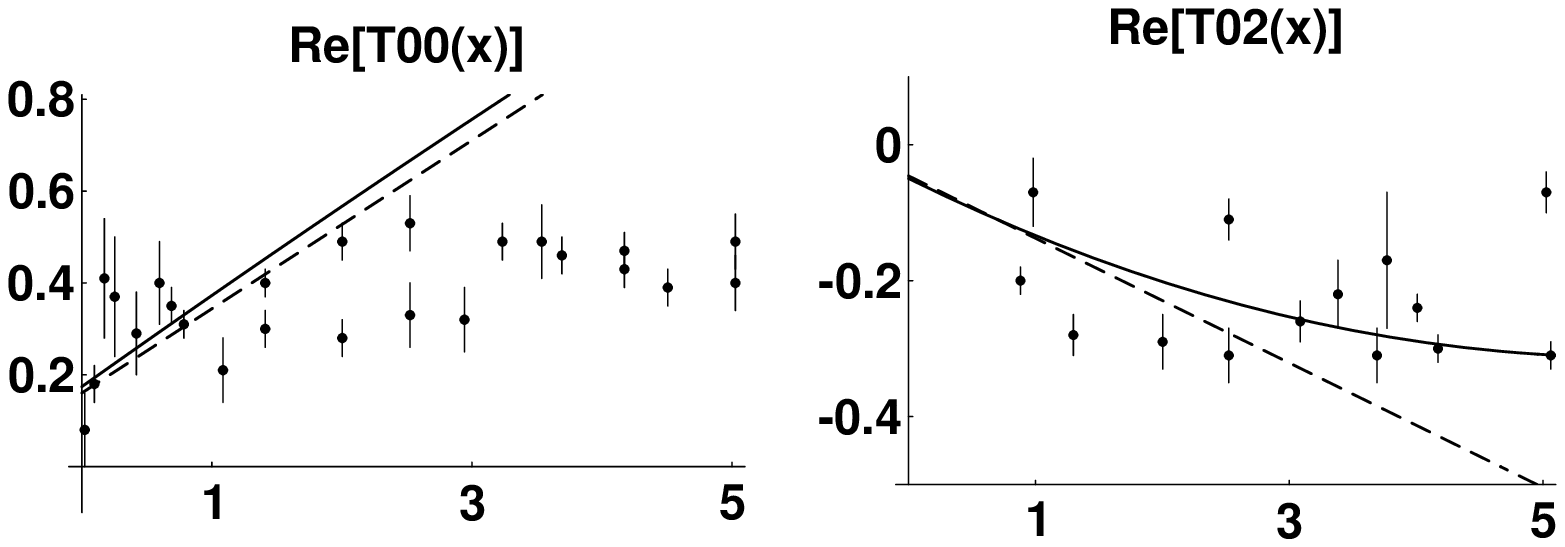,rheight=7.3cm,height=12.0cm} }
\caption{Partial wave amplitudes in $\pi$-$\pi$ scattering
($x= E^2/(4 m_\pi^2) - 1$): solid line - this calculation; dashed line -
current algebra.\protect\cite{SW66} \label{mpwaFQ}}
\end{figure}

\begin{figure}[htb] 
 \centering{\
  \epsfig{figure=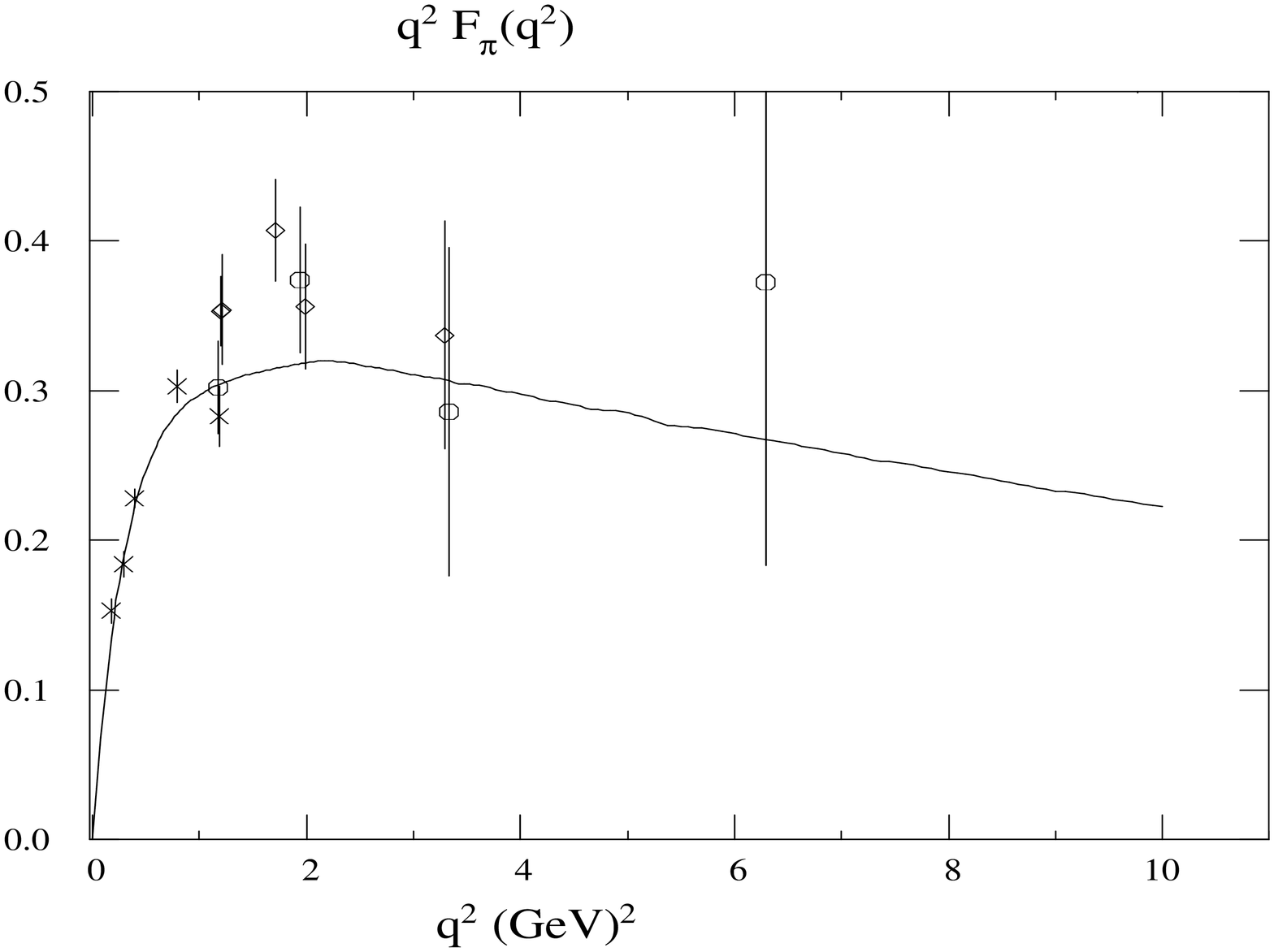,rheight=6.0cm,height=6.0cm,angle=-90} }
\caption{Calculated form of $F_\pi(q^2)$. The experimental data are from
Refs.\protect\citenum{Exp73} (crosses), \protect\citenum{Exp76} (diamonds) and
\protect\citenum{Exp78} (circles).\label{FQQplot}}
\vspace*{-\baselineskip}
\end{figure}

The low-energy physical observables calculated with this parameter set are
compared with their physical values in Table~1, where
\mbox{$m^{\rm ave}=\left( m_u +m_d\right)/2$ } and the ``experimental'' value
of
\mbox{$\langle \bar q q \rangle^{\frac{1}{3}}$} is that typically used in QCD
sum rules analysis.  The calculated quantities were evaluated at the listed
value of $m_\pi$; i.e., the chiral limit expressions were not used, but
the corrections are $<1$\% in each case.  The agreement is excellent.
\begin{table}[htb]
\begin{center}
\begin{tabular}{|c|l|l|} \hline
   & Calculated  & Experiment  \\ \hline
  $f_{\pi} \; $    &  ~0.0839 GeV &   ~0.0931 $\pm$ 0.001     \\ \hline
  $-\langle \bar q q \rangle^{\frac{1}{3}}_{1\,{\rm GeV}^2}$ & ~0.211 &
        ~0.220 $\pm$ 0.01\\ \hline
  $m^{\rm ave}_{1\,{\rm GeV}^2}$ & ~0.0061 & ~0.0075 $\pm$ 0.004
                \\ \hline
 $m_{\pi} \; $    & ~0.127  & ~0.138  \\ \hline\hline
 $r_\pi \;$ & ~0.596 fm & ~0.663$\pm$ 0.006  \\  \hline\hline
 $g_{\pi^0\gamma\gamma}\;$ & ~0.497 (dimensionless) & ~0.504 $\pm$ 0.019\\
                        \hline\hline
 $a_0^0 \;  $ & ~0.174  & ~0.21$\pm$0.02 \\ \hline
 $a_0^2 \;  $ & -0.0496 & -0.040 $\pm$ 0.003 \\ \hline
 $a_1^1 \;  $ & ~0.0307 & ~0.038 $\pm$ 0.002\\ \hline
 $a_2^0 \; $  & ~0.00161 & ~0.0017 $\pm$ 0.0003\\ \hline
 $a_2^2 \;  $ & -0.000251 & \\ \hline
\end{tabular}
\caption{\label{tableone}Comparison of {\it tree-level} DSE calculations with
experiment.}
\vspace*{-5mm}
\end{center}
\end{table}
The $\pi$-$\pi$ partial wave amplitudes associated with $a_0^0$ and $a_0^2$,
calculated using the formulae in Ref.~\citenum{RCSI94}, are plotted in
Fig.~\ref{mpwaFQ} and can be seen to be in agreement with the data up to
$x\approx 3$, which corresponds to $E\approx 4 m_\pi$.  The same is true of
the higher partial wave amplitudes.\cite{CDRpion}
The calculated form of $F_\pi(q^2)$ is presented in Fig.~\ref{FQQplot} and,
given that the extraction of the ``experimental'' point at $q^2=6.3$~GeV$^2$,
measured in pion electroproduction\cite{Exp78} is strongly model dependent,
the agreement with the experimental data is again excellent.
\subsection{Concluding Remarks}
\label{sec_four}
\setcounter{equation}{0}
In principle, QCD can be solved using the Dyson-Schwinger equation (DSE)
framework, however, in practice, obstacles to achieving this goal remain.
The successes and remaining challenges are discussed in detail in
Ref.~\citenum{DSErev}.

Herein, the practical application of the DSE approach to the calculation of
physical observables is illustrated.  By its very nature, the approach
directly incorporates all of the known large spacelike-$q^2$ behaviour of
QCD.  It has a phenomenological, model dependent aspect that is tied to the
fact that the gluon propagator, $D_{\mu\nu}(q)$, is unknown for
spacelike-$q^2<1$~GeV$^2$.  This has the benefit that, in calculating
experimental observables in this approach, one obtains a representation of
these observables in terms of the infrared structure of $D_{\mu\nu}(q)$ and
can thereby use precision experimental measurements to determine the infrared
form of the gluon propagator.

In the DSE approach the pion has an intrinsic size; i.e., it is not
pointlike, and its dominant determining characteristic is its dressed quark
core, which is described by the proper pion-quark vertex function in
\Eq{Gpi}.  The calculations reported herein are {\it tree-level}
calculations, which here actually means that they are obtained with the
minimal number of dressed-quark loops.  The agreement between theory and
experiment is at the 10\% level, which is consistent with
Refs.~\citenum{HRM,ABR} where {\it nonpointlike} $\pi$-loop contributions are
shown to provide $<15$\% corrections.
\begin{table}[htb]
\vspace*{-\baselineskip}
\begin{center}
\begin{tabular}{|c|r|r|}\hline
Low Energy Constant & {\it tree-level}\ DSE &
1-$\rule{0mm}{10pt}\pi^{\rm point}$-loop ChPT\protect\cite{GL84} \\ \hline
$\rule{0mm}{10pt}\bar\ell_1 $ & $-6.1$ & $-2.3 \pm 3.7$ \\  \hline
$\rule{0mm}{10pt}\bar\ell_2 $ & $~6.6$ & $~6.0 \pm 1.3$ \\  \hline
$\rule{0mm}{10pt}\bar\ell_6$  & $11.7$ & $16.5 \pm 1.1$ \\ \hline
\end{tabular}
\caption{\label{LECs}Low-energy constants of ChPT inferred from {\it
tree-level} DSE calculations of experimental observables.}
\vspace*{-\baselineskip}
\end{center}
\end{table}

Clearly, experimental observables can be calculated directly in the DSE
approach.  From these calculations, however, one might infer values of the
low-energy constants used in Chiral Perturbation Theory (ChPT) to parametrise
the solution of the chiral Ward identities in QCD.  This will provide
implicit, nonlinear relations between these constants and the few, underlying
parameters of QCD.  The results inferred for $\bar \ell_1$, $\bar
\ell_2$ and $\bar \ell_6$ from DSE-{\it tree-level} calculations are
presented in Table.~\ref{LECs} and compared with the 1-point-pion-loop
corrected values used in ChPT.  (The 1-{\it nonpointlike}-pion-loop
correction to $r_\pi$ estimated in Ref.~\citenum{ABR} yields
$\bar\ell_6\simeq 13.0-13.5$.)  This comparison indicates that {\it
tree-level} DSE calculations incorporate those effects of point-pion-loops in
ChPT that serve merely to mock-up the pion's internal structure.
Calculations\nopagebreak[4] from\nopagebreak[4] which\nopagebreak[4] the
value of other of these constants can be inferred are underway.\cite{ICDRT}

\pagebreak

\hspace*{-\parindent}{\it Acknowledgments.} I would like to thank and
congratulate the organisers, Aron Bernstein and Barry Holstein, for bringing
about this timely and stimulating meeting, and the workshop secretary, Joanne
Gregory, for ensuring that it ran smoothly.  This work was supported by the
Department of Energy, Nuclear Physics Division under contract number
W-31-109-ENG-38.



\end{document}